\documentstyle[prb,aps,preprint]{revtex}
\begin{document}
\title{Large $N$ study of extreme type II superconductors in a 
magnetic field}
\author{Claude de Calan}
\address{Centre de Physique Th\'eorique,  
Ecole Polytechnique, 
F-91128 Palaiseau Cedex, France}
\author{Adolfo P. C. Malbouisson}
\address{Centro Brasileiro de Pesquisas F\'{\i}sicas-CBPF, Rua 
Dr. Xavier Sigaud 150, Rio de Janeiro, RJ 22290-180, Brazil}
\author{Flavio S. Nogueira}
\address{Institut f\"ur Theoretische Physik, 
Freie Universit\"at Berlin, Arnimallee 14, D-14195 Berlin, Germany}

\date{Received \today}

\maketitle

\begin{abstract}
The large $N$ analysis of an extreme type II superconductor is 
revisited. It is found that the phase transition is of 
second-order in dimensions $4<d<6$. For the physical dimension 
$d=3$ no sign of phase transition is found, contrary to 
early claims.         
\end{abstract}
\draft

High temperature superconductors have a very large Ginzburg 
parameter, typically $\kappa\sim 100$. For this reason, it  
seems to be a good approximation to neglect magnetic thermal 
fluctuations in the Ginzburg-Landau (GL) model. For 
$\kappa\gg 1$ the Hamiltonian density of the 
GL model in an external magnetic field is written as 

\begin{equation}
\label{GL}
H=|(\nabla-ie{\bf A})\phi|^2+\frac{u}{2}\left(|\phi|^2+\frac{m^2}{u}
\right)^2,
\end{equation}
where $\nabla\times{\bf A}={\bf H}$ and $m^2=a(T-T_c)$ with 
$a>0$. This model Hamiltonian describes superconductors in the 
extreme type II limit. 

Early renormalization group calculations 
performed by Br\'ezin {\it et al.} \cite{Brezin} using the model (\ref{GL}) 
indicated that the phase transition is of first-order. 
This result 
has been obtained in the lowest Landau level (LLL) approximation with 
an $\epsilon=6-d$-expansion. Later, 
Affleck and Br\'ezin \cite{Affleck} carried a large $N$ calculation and 
have obtained also a first-order phase transition. 
The situation seems to be different from the Halperin {\it et al.} 
calculation \cite{HLM} in zero field but with magnetic fluctuations.  
There, the $\epsilon=4-d$-expansion leads to a first-order transition 
while a large $N$ analysis gives a second-order transition. 

A large $N$ analysis performed by Radzihovsky \cite{Radz} leads to an 
opposite conclusion to that of Affleck and Br\'ezin \cite{Affleck}. 
This author obtained instead that the transition is of second-order 
in dimensions $4<d<6$. 

In order to solve the controversy, 
in this paper we revisit the large $N$ problem by performing a simpler 
analysis with respect to the previous ones.        
We will see that the leading order is just the 
Hartree approximation considered by Lawrie \cite{Lawrie} in his study 
of the scaling in high-$T_c$ superconductors. 
Our main results are the following.  
We find a second-order phase transition in $4<d<6$, which agree 
with Radzihovsky \cite{Radz}, disagreeing with Affleck and Br\'ezin 
\cite{Affleck}. We have computed the 
effective interaction at large $N$ in $4<d<6$ and obtained 
the $\beta$-function. We find an infrared stable fixed point, 
consistent with the second-order transition scenario. 
In three dimensions we are not able to find a phase transition 
in the context of the large $N$ approximation.

In the following we will assume that the external magnetic field is 
parallel to the $z$ axis and that the gauge ${\bf A}=(0,xH,0)$ has 
been chosen. We will consider the model (\ref{GL}) with $N$ complex 
components and take the large $N$ limit at $Nu$ fixed. 
In order to treat the large $N$ limit, we will introduce an auxiliary field 
$\sigma$ and obtain the transformed Hamiltonian:

\begin{equation} 
\label{GLHS} 
H'=|(\nabla-ie{\bf A})\phi|^2+i\sigma\left(|\phi|^2+\frac{m^2}{u}\right) 
+\frac{1}{2u}\sigma^2. 
\end{equation}
The new Hamiltonian $H'$ is Gaussian in $\phi$. This allows a straightforward  
integration of $\phi$ to obtain the following effective action:

\begin{equation} 
\label{Seff} 
S_{eff}=N Tr\ln(-\partial^2+2i\omega x\partial_y+\omega^2 x^2+i\sigma) 
+\int d^3 r\left(\frac{m^2}{u}i\sigma+\frac{1}{2u}\sigma^2\right), 
\end{equation}
where $\omega=eH$ is the cyclotron frequency. 
The leading order in $1/N$ is obtained through the minimization of 
$S_{eff}$ with respect to $\sigma$. We will take $\sigma$ as being 
uniform and given by $\sigma=-i\sigma_0$. In this way we can 
easily evaluate the trace of the logarithm in (\ref{Seff}) using the 
eingevalues of the operator 
$-\partial^2+2i\omega x\partial_y+\omega^2 x^2+\sigma_0$, which are the 
well known Landau levels. Close to the critical point,  
the most relevant of the Landau levels is the lowest one. 
By doing the minimization of (\ref{Seff}) taking only the LLL 
simplifies considerably the calculation. The field $\phi$ should be 
written in terms of the Landau level basis as follows:

\begin{equation}
\phi({\bf r})=\sum_n\int\frac{dp_y}{2\pi}\int\frac{dp_z}{2\pi}
\hat{\phi}_{n,p_y,p_z}\chi_{n,p_y,p_z}({\bf r}),
\end{equation}
where $\chi_{n,p_y,p_z}({\bf r})$ are the Landau level eingenfunctions given 
by

\begin{equation}
\chi_{n,p_y,p_z}({\bf r})=\frac{1}{\sqrt{2^n}n!}\left(\frac{\omega}{
\pi}\right)^{1/4}e^{i(p_z z+p_y y)}e^{-\omega(x-p_y/\omega)^2/2}
H_n\left(\sqrt{\omega}x-\frac{p_y}{\sqrt{\omega}}\right),
\end{equation}
with energy eigenvalues 
$E_{n}(p_{z})=p_{z}^{2}+(2n+1)\omega+\sigma_0$
and where $H_n$ are the Hermite polynomials.

If 
all the Landau levels are retained, the model is renormalizable 
in dimensions $2<d\leq 4$, just as the case of pure $\phi^4$ 
theory. However, close to the critical field $H_{c_2}$, the 
LLL correspond to the most important fluctuations and the 
upper and lower critical dimensionnalities are changed. 
Let us see how this happens in the large $N$ limit. 
To this end, we consider $p_z$ as a $(d-2)$-dimensional 
vector. By minimizing Eq. (\ref{Seff}) with respect to 
$\sigma_0$ retaining only the LLL, 
we obtain the following gap equation:

\begin{equation}
\label{gapd}
\sigma_0=m^2+\frac{Nu\omega}{2\pi^2}\int\frac{d^{d-2}p_z}{(2
\pi)^{d-2}}\frac{1}{p_z^2+\omega+\sigma_0}.
\end{equation}
The critical point correspond to $\sigma_0=-\omega$ and is 
attained for $m^2=m^2_c$. Therefore, 

\begin{equation}
\label{gapdCP}
0=m_c^2+\omega+\frac{Nu\omega}{2\pi^2}\int\frac{d^{d-2}p_z}{(2
\pi)^{d-2}}\frac{1}{p_z^2}.
\end{equation}
From Eq. (\ref{gapdCP}) we see that 
the lower critical dimension is $d=4$. Combining Eqs. 
(\ref{gapd}) and (\ref{gapdCP}) we obtain 

\begin{equation}
\label{gapd1}
\sigma_0+\omega=m^2-m_c^2(\omega)-\frac{Nu\omega}{2\pi^2}
\int\frac{d^{d-2}p_z}{(2\pi)^{d-2}}\frac{\sigma_0+\omega}{
p_z^2(p_z^2+\omega+\sigma_0)}.
\end{equation}
From the above equation we see that the upper critical 
dimension is $d=6$. 
As $\sigma_0+\omega\to 0$, the integral in Eq. (\ref{gapd1}) 
dominates and behaves as $\sim(\sigma_0+\omega)^{(d-4)/2}$. 
We obtain therefore the scaling relation  

\begin{equation}
\label{scaling}
\sigma_0+\omega\sim [m^2-m_c^2(\omega)]^{2\nu}\sim[T-T_c(\omega)]^{2\nu}, 
\end{equation}
where the critical exponent 
$\nu=1/(d-4)$. This value of $\nu$ is exact at large $N$.   
   
Let us calculate the quadratic fluctuations in $\sigma$. This will 
allow us to obtain the $\sigma$ propagator which corresponds 
to the effective $|\phi|^4$ coupling. In order to perform this 
calculation, we will substitute in Eq. (\ref{Seff})  
$i\sigma=\sigma_0+i\delta\sigma$, where $i\delta\sigma$ is 
a small fluctuation around $\sigma_0$. Thus, up to quadratic 
order in $\delta\sigma$, the effective action is 

\begin{equation}
\label{fluc}
S_{eff}=S_{eff}^{(0)}+\frac{1}{2u}\int d^3 r\int d^3 r'
[\delta^3({\bf r}-{\bf r}')+Nug_0({\bf r},{\bf r}')g_0({\bf r}',{\bf r})]
\delta\sigma({\bf r})\delta\sigma({\bf r}')+(h.o.t.), 
\end{equation}
where $S_{eff}^{(0)}$ corresponds to the saddle point 
contribution and 
$g_0({\bf r},{\bf r}')$ is the Green function of the operator 
$-\partial^2+2i\omega x\partial_y+\omega^2 x^2+\sigma_0$. 
In this formalism, the study of the $|\phi|^4$ coupling is 
replaced by the study of the $\sigma$-propagator, which allows the 
definition of a unique coupling constant at the LLL. This situation 
contrasts with the $\epsilon=6-d$ expansion\cite{Brezin}. 
The LLL analysis of the $\sigma$-propagator gives 
the effective $|\phi|^4$ interaction in momentum space 
and at the critical point,

\begin{equation}
\label{interac}
U_\sigma({\bf p})=
\frac{u}{1+Nu\omega c(d)
e^{-\frac{1}{2\omega}(p_x^2+p_y^2)}(p_z^2)^{\frac{d-6}{2}}},
\end{equation}
where $c(d)=c_0 2^{2-d}\pi^{1-d/2}\Gamma(1-d/2)\Gamma^2(d/2-2)
/\Gamma(d-4)$. 

If we take as a running scale $\mu=|p_z|$, we can define 
the dimensionless coupling 

\begin{equation}
g=U_\sigma(p_x=0,p_y=0,\mu)\omega\mu^{d-6}.
\end{equation}
It is then straightforward to obtain the $\beta$-function 
$\beta(g)\equiv\mu\partial g/\partial\mu$:

\begin{equation}
\label{beta46}
\beta(g)=(6-d)[-g+Nc(d)g^2],
\end{equation}
with an infrared stable fixed point 

\begin{equation}
g_*=\frac{1}{Nc(d)}.
\end{equation}
Therefore, in $4<d<6$ our analysis agrees with the one by 
Radzihovsky\cite{Radz}, pointing to a second-order phase transition. 
It disagrees, however, with the analysis by 
Affleck and Br\'ezin\cite{Affleck}, 
who obtained that the transition is of first-order 
in $4<d<6$. 

The question now is what happens at $d=3$, which is the physical 
dimensionality of the superconductor. Since the lower critical 
dimension is found to be four, we may expect that severe 
infrared singularities be present for $2<d<4$. Let us confine 
ourselves to $d=3$ and study the gap equation and 
coupling constant. 

For $d=3$ the integral in (\ref{gapd}) is ultraviolet 
convergent. By performing explicitly the integration, we 
obtain

\begin{equation}
\label{gap3}
\sigma_0+\omega=m^2+\omega+
\frac{Nu\omega}{4\pi^2}\frac{1}{\sqrt{\sigma_0+\omega}}.
\end{equation} 
We see that it is not possible to make 
$\sigma_0+\omega\to 0$ as in $4<d<6$. This means that if there 
exists a phase transtion in $d=3$, it is surely not a 
second-order one. 
We have computed the free energy and the specific heat, which 
shows no discontinuity or peak. 
Thus, we are not able to find any 
sign of phase transition in $d=3$ in this large $N$ calculation. 
This is in 
contrast with experiments \cite{Experiments}  
pointing to the existence of a 
first-order phase transition. There are also 
theoretical arguments \cite{Russians} for the melting of the vortex lattice 
that give also a first-order phase transition. 
The absence of such a transition in our study could be a specific 
feature of the $N\to\infty$ case. Note that, once more, our 
analysis disagrees with the one by Affleck and Br\'ezin\cite{Affleck}, 
since these authors conclude that there is a first-order phase 
transition in $d=3$.

Summarizing, we have performed a large $N$ study of an extreme 
type II superconductor. The second-order transition in 
$4<d<6$ found by Radzihovsky\cite{Radz} years ago is 
recovered using more simple arguments. Our analysis disagrees 
with the one by Affleck and Br\'ezin\cite{Affleck}, where 
instead a first-order phase transition is found. In $d=3$ we 
are not able to find any sign of phase transition. It must be 
kept in mind, however, that at low $N$ the results can be 
very different. Indeed, other theoretical analysis at 
$d=3$ and $N=2$ seem to indicate that the phase diagram is richer 
\cite{Tesanovic}. Monte Carlo simulations 
\cite{Nguyen} indicate that there is a tricritical point 
in the $H-T$ phase diagram, allowing for the possibility of 
a first- or second-order phase transitions. This tricritical 
point is reminiscent from the tricritical point obtained in 
the duality scenario for the zero-field superconductor\cite{Kleinert}. 
The point of view of Refs. \onlinecite{Tesanovic} and 
\onlinecite{Nguyen} seem to be confirmed by recent experiments 
\cite{Marcenat}.

{\bf Acnowledgements}

APCM would like to thank the hospitality of the Centre de Physique 
Th\'eorique of the Ecole Polytechnique, where part of this work has 
been done. 
CC and FSN would like to thank the hospitality of the Centro Brasileiro 
de Pesquisas F\'{\i}sicas. The authors acknowledge financial 
support from the brazilian agencies CNPq and FAPERJ.  
The work of FSN is supported by the 
Alexander von Humboldt foundation.

\end{document}